\documentclass[preprint,12pt]{elsarticle}

\usepackage{amssymb}

\usepackage{lineno}

\usepackage{ulem} 
\usepackage{color} 

\biboptions{sort&compress}

\journal{Nuclear Inst. and Methods in Physics Research, A}

\begin{document}

\begin{frontmatter}



\title{Surface detectors of the TAx4 experiment}


\author[1]{R.U.~Abbasi}
\author[2]{M.~Abe}
\author[1,3]{T.~Abu-Zayyad}
\author[3]{M.~Allen}
\author[4]{Y.~Arai}
\author[3]{E.~Barcikowski}
\author[3]{J.W.~Belz}
\author[3]{D.R.~Bergman}
\author[3]{S.A.~Blake}
\author[3]{R.~Cady}
% ...

\address[1]{Department of Physics, Loyola University Chicago, Chicago, Illinois, USA}
\address[2]{The Graduate School of Science and Engineering, Saitama University, Saitama, Saitama, Japan}
\address[3]{High Energy Astrophysics Institute and Department of Physics and Astronomy, University of Utah, Salt Lake City, Utah, USA}
\address[4]{Graduate School of Science, Osaka City University, Osaka, Osaka, Japan}
% ...

\cortext[cor1]{Corresponding author.}
\fntext[fn1]{Deceased.}



\begin{abstract}
Telescope Array (TA) is the largest ultrahigh energy cosmic-ray (UHECR) observatory in the Northern Hemisphere. It explores the origin of UHECRs by measuring their energy spectrum, arrival-direction distribution, and mass composition using a surface detector (SD) array covering approximately 700 km$^2$ and fluorescence detector (FD) stations. TA has found evidence for a cluster of cosmic rays with energies greater than 57 EeV. 
In order to confirm this evidence with more data, it is necessary to increase the data collection rate.
We have begun building an expansion of TA that we call TAx4. In this paper, we explain the motivation, design, technical features, and expected performance of the TAx4 SD. 
We also present TAx4’s current status and examples of the data that have already been collected.
\end{abstract}


\end{frontmatter}



\section{Introduction}
\label{intro}

\begin{figure}
      \centering
      \includegraphics[width=0.5\textwidth]{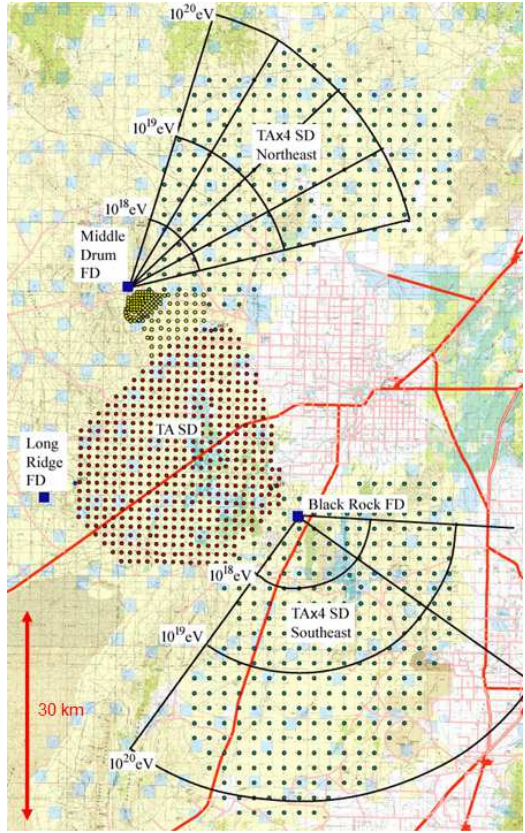}
      \caption{Map of the Telescope Array (TA) site. Each green circle in the northeast and southeast corresponds to the planned location of a TAx4 surface detector (SD); the spacing of the TAx4 SDs is 2.08 km. Each red circle in the west shows the location of a TA SD; the spacing of these is 1.2 km. 
      Each yellow circle shows the location of an SD belonging to the low-energy TALE experiment~\cite{TALE}. 
      The two fan shapes drawn with black lines are the fields of view of the TAx4 fluorescence detectors (FDs). 
      Four telescopes have been built in the northern Middle Drum FD station, and eight telescopes in the southern Black Rock FD station. 
      The overlap of the locations of the SDs and the fields of view of the FDs enables SD-FD hybrid observation.}
      \label{map} 
\end{figure}

Telescope Array (TA) is the largest cosmic-ray observatory in the Northern Hemisphere~\cite{TA}. The aim of TA is to explore the origin and nature of ultrahigh energy cosmic rays (UHECRs) above 10$^{18}$ eV.
The TA experiment has 507 plastic scintillator surface detectors (SDs) arrayed in a square grid with 1.2 km spacing covering approximately 700 km$^{2}$.
TA also has three fluorescence detector (FD) stations, two of which have 12 telescopes, while the third has 14 ~\cite{TAFD1, TAFD2, TAFDMD}. Full operation began on May 11, 2008.
The duty cycle of the SD array has been 95\% throughout the 11-year observation period on average. The FD duty cycle has been approximately 10\%, because the data is taken only on clear moonless nights.
The layout of TA is shown in Fig.~\ref{map}; its central laser facility (CLF) is located at 39.30$^{\circ}$N, 112.91$^{\circ}$W and 1370 m above sea level in Utah, USA.

In previous work using TA SD data collected over five years, we found that 19 out of 72 observed cosmic rays with energies above 57 EeV were concentrated within a single 20$^\circ$-radius circle (the ``hotspot'')~\cite{hotspot}. 
The result had a maximum local statistical significance of 5.1$\sigma$; for the hotspot to arise by chance anywhere in the field of view would require a 3.4$\sigma$ fluctuation, which is the global statistical significance of the result.

Several additional indications of cosmic-ray energy-spectrum anisotropy have also recently been observed by the TA experiment \cite{Decl, SpecAniso}.
A difference in the spectral cutoffs above and below the declination $\delta = 24.8^{\circ}$ has been reported in~\cite{Decl}:
the high and low declination band cutoffs were found to be $10^{19.85^{+0.03}_{-0.03}}$ eV and $10^{19.59^{+0.05}_{-0.07}}$ eV, respectively.
The difference was updated using the latest data in~\cite{Decl2}.
The statistical significance of the difference of the break points of the energy spectra was estimated to be 4.3$\sigma$.
A relative energy distribution test was also done comparing events inside oversampled spherical caps of equal exposure to those outside.
The center of maximum significance has been found at right ascension $\alpha =$ 9$^h$16$^m$ and declination $\delta =$ 45$^{\circ}$, and has
an excess of arrival directions for particles with energies above $10^{19.75}$ eV and a deficit for $10^{19.2}$ eV $\le E < 10^{19.75}$ eV~\cite{SpecAniso}.
The post-trial significance was estimated to be 3.7$\sigma$.
Correlation of arrival directions with energies was also reported in~\cite{Wedge};
the post-trial significance of the correlation in this case was estimated to be 4.2$\sigma$.
These results were obtained using seven to eleven years of observations with the TA SD.

In order to accelerate the pace of data collection as we further investigate these apparent departures from isotropy, we have developed a quadrupled TA detector. We call this proposed larger detector ``TAx4.''

The SD array of the TAx4 experiment is designed expressly to study cosmic rays with energies above 57 EeV. The spacing of the TAx4 SD array is 2.08 km, in contrast to the TA SD’s 1.2 km. 
The area covered by the TAx4 SDs over the area covered by the TA SDs can be estimated to be (2.08/1.2)$^2 \sim 3$ if there are large number of SDs.
The additional 500 TAx4 SDs cover approximately three times more area than the 507 TA SDs; the combined coverage of the TAx4 and TA SDs is approximately 2800 km$^2$.
Fig.~\ref{map} shows the planned locations of the TAx4 detectors. Two FD stations for the TAx4 experiment have also been designed. The field of view of each FD covers the sky above the area of the SDs so that cosmic-ray events can be observed simultaneously by FD and SD; we call such events “hybrids.”

The design and the expected performance (based on simulations) of the TAx4 SDs are shown in the next section.
The status of SD construction is discussed in Section \ref{Construction}. 
The results of calibration and quality checks are given in 
Section \ref{Calibration}.
The data acquisition system is shown in 
Section \ref{DAQ}.
Section \ref{FuturePros} provides a summary.

\section{Design and performance}
\label{Design}

\begin{figure}
      \centering
      \includegraphics[width=0.5\textwidth]{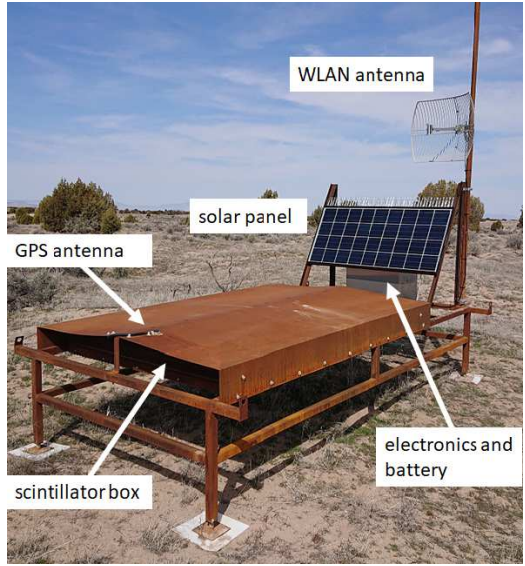}
      \caption{Photograph of one of the deployed TAx4 SDs. 
        The scintillator box is 
        shaded 
        by a roof. 
        The wireless LAN antenna is fixed to the antenna pole aiming at the communication tower. 
        The solar panel facing to south charges the battery in the stainless-steel box and supplies power to the electronics and other devices.
        Bird spikes are attached on the support frame of the solar panel for each TAx4 SD.
        The open spaces on the northern and southern sides of the scintillator box under the roof are covered with mesh plates for each TAx4 SD in addition to the cover plates of the detector sides. 
      }
      \label{SDpicture} 
\end{figure}

The basic design of a TAx4 SD is the same as that of a TA SD~\cite{TANIM}.
The most essential component of an SD unit is a pair of plastic scintillator layers, each of 1.2 cm thickness and 3 m$^2$ area.
Each plastic scintillator layer is composed of eight plastic scintillator sheets. The length, width and thickness of each sheet are 150 cm, 25 cm and 1.2 cm, respectively.
The scintillation light from each layer is collected by wavelength shifting (WLS) fibers and a photomultiplier tube (PMT).
Coincidence signals from the two layers of scintillators are taken to measure the single-muon peak 
clearly removing random noise with low energy from each layer
for the calibration of signals.
The data are taken with the SD electronics mounted on each SD.
The SD electronics consists of a motherboard, a wireless LAN modem and a charge controller.
The motherboard includes electronics of readout of signals and trigger generation and a GPS receiver.
The data taken at each SD are collected by the host electronics at the corresponding communication tower using 2.4 GHz wireless LAN communication.
There are six communication towers, and each communication tower collects the data from the SDs in each sub-array. The distance between the communication tower and the SD in a sub-array is smaller than 19 km~\cite{comtower}.
Fig.~\ref{SDpicture} shows a typical TAx4 SD unit.
Bird spikes, plates under the roof, and cable-protector tubes have been added to the TAx4 SDs to avoid damage caused by small animals in the field.

\begin{figure}
      \centering
      \includegraphics[width=0.5\textwidth]{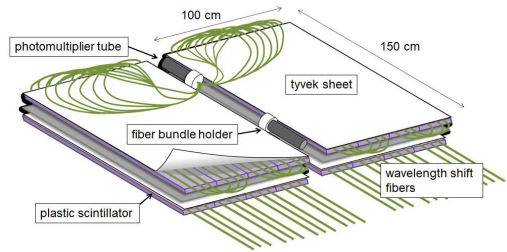}
      \caption{
        Schematic view inside the scintillator box. The plastic scintillators 
        with 
        wavelength-shifting 
        (WLS)
        fibers attached 
        to the grooves on the scintillator surfaces are covered
        with 
        the two Tyvek sheets.
        The length of each fiber is 6.1 m and the horizontal fiber spacing is 4 cm.
        The 28 fibers in the same layer are folded at the layer’s edge, and 56 fiber edges are bundled.
        The fiber bundle edge is polished and connected to the surface of the photomultiplier tube (PMT) with optical grease. The WLS fibers on the lower layer 
        are drawn as straight lines in this figure,
        but they are actually folded, bundled and attached to the edge of the other PMT 
        in the same way as those on the upper layer.
      }
      \label{ScintiBox} 
\end{figure}

Fig.~\ref{ScintiBox} shows a schematic view of the interior of the scintillator box.
We use R8619 Hamamatsu PMTs~\cite{R8619} in the TAx4 SDs; these have a higher quantum efficiency 
than those used in the original TA SDs.
We use the same WLS fibers (Y-11 Kuraray) as the TA SDs but change their arrangement for cost reduction:
the fiber spacing is enlarged from 2 cm to 4 cm, the length of the fibers is extended from 5 m to 6.1 m, and the number of fibers is reduced from 104 to 28 on each layer.
The total length of the WLS fibers is reduced by 67\% from that of the TA SD~\cite{TANIM}.
The material of the plastic holder of the WLS fiber bundle 
attached to the PMT surface
has also been changed from transparent acrylic to white polyacetal resin in order to reduce losses of photons,
and the arrangement of the fiber bundle in the holder has been modified to use only the center of the photocathode of the PMTs.
The change of the material of the plastic holder
increased the number of photons incident on 
the PMTs by 11\%~\cite{Saito}.
As a result of the modifications inside of the scintillator box, the average number of photoelectrons from the PMTs of the TAx4 SDs is similar to that of the TA SDs. The number of photoelectrons will be shown in detail in Section~\ref{Calibration}.
For the TA SD, we have attached the PMTs at the bottom of each scintillator box during the final assembly after the boxes arrived in Utah.
In the case of the TAx4 SD, we have transported the boxes from Japan with the PMTs already attached, which reduced the final assembly workload.
Also, for the TAx4 SD, we have attached the PMTs using more solid metal fitting than in the case of the TA SD in order to avoid damaging the PMTs during the transportation.

\begin{figure}
  \centering
  \includegraphics[width=.5\textwidth]{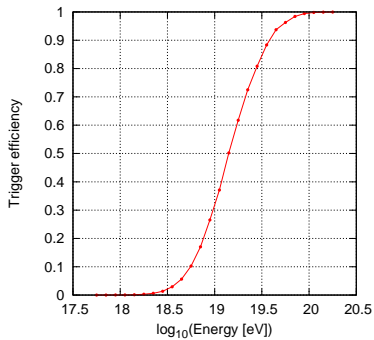}
  \caption{
    Trigger efficiency of the TAx4 surface detectors for 
    proton cosmic-ray air showers
    as a function of the primary energy.
  }
  \label{TrigEff} 
\end{figure}

The performance of the TAx4 SD array was estimated using Monte Carlo (MC) simulations of air showers and of detectors, similar to those already done for the TA SDs~\cite{Ivanov}. 
Fig.~\ref{TrigEff} shows the simulated energy dependence of the trigger efficiency.
The simulations used a 10$\times$10 SD array on a square grid with 2.08 km spacing, with an assumed 100\% efficiency of detectors and data acquisition.
The detailed settings for the air-shower simulations were the same as those in~\cite{TASDMC} for the TA SDs:
cosmic-ray protons were simulated using QGSjet-II-03~\cite{QGSjet2} high-energy hadron interaction model in the region of zenith angles between 0$^{\circ}$ and 60$^{\circ}$. 
The detector-simulation conditions were the same as those in the TA SD simulations except for the trigger condition: the gate width for the trigger of the air showers was enlarged from 8 $\mu$s for the TA SDs to 14 $\mu$s for the TAx4 SDs considering larger spacing. 
The conversion from the zenith angle and S800 (the signal density at 800 m from the shower axis) to the primary energy was done in the same manner as in the TA SD analysis~\cite{TASDEspectrum}.
For comparison with the energies measured calorimetrically by the FDs, the calculated primary energies were rescaled by 1/1.27~\cite{TASDEspectrum} (a factor determined empirically by examining SD-FD hybrid events observed by TA).

We obtained 25\% energy resolution, 2.2-degree angular resolution, and 95\% reconstruction efficiency of cosmic rays with energies above 57 EeV. 
Both energy and angular resolution of the TAx4 SDs are worse than those of the TA SDs due to the wider spacing, but the resolution is good enough for the anisotropy studies.
Table \ref{t1} compares the performance of the TAx4 SD with that of the TA SD. 

\begin{table}[tbh]
\caption{Simulated performances of TAx4 and TA surface detector (SD) arrays for 
cosmic rays
with E $>$ 57 EeV}
\label{t1}
\begin{tabular}{cc}
\hline
SD Array & Angular Resolution (degrees) \\
TA SD & 1.0 \\
TAx4 SD & 2.2 \\
\hline
SD Array &  Energy Resolution (\%) \\
TA SD & 15 \\
TAx4 SD & 25 \\
\hline
SD Array & Reconstruction Efficiency (\%) \\
TA SD & 99  \\
TAx4 SD & 95  \\
\hline
\end{tabular}
\end{table}

\section{Construction}
\label{Construction}
 
\begin{figure}
  \centering
  \includegraphics[width=.4\textwidth]{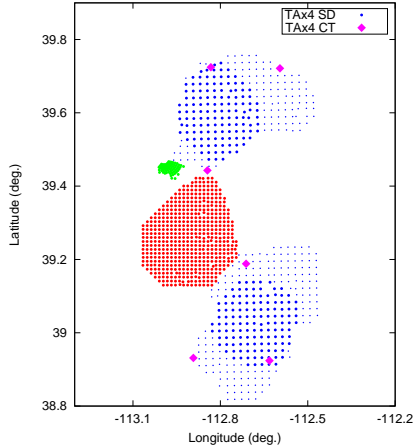}
  \caption{Locations of deployed surface detectors (SDs) and constructed communication towers. 
  The blue circles are the locations of deployed SDs.
  The blue dots are planned locations of future SDs deployments. 
  The pink diamonds are the locations of communication towers.
  The red circles show the locations of the TA SDs.
  The green circles show the locations of the TALE experiment’s SDs.
  }
  \label{DeployedSDs} 
\end{figure}

The TAx4 SD scintillator boxes were assembled mostly in Japan, starting in 2015 (30 were assembled in Korea in 2018) and then transported to the USA.
Final assembly of the SDs was performed at the Cosmic-Ray Center in Delta City, Utah.
A total of 257 assembled SDs were deployed in February and March 2019. 
Fig.~\ref{DeployedSDs} shows the locations of the deployed SDs. 
The arrangement of the 257 SDs was meant to optimize the number of hybrid events above 10 EeV, taking into account practical considerations of radio contact between the communication towers and the six sub-arrays of the SD array. 
We visited all locations by buggy or helicopter in advance to investigate the suitability of the deployment and line-of-sight to the communication towers. 
The assembled SDs were transported from the Cosmic-Ray Center to the staging areas near the TAx4 sites using flatbed trucks.
All SDs were deployed from the staging areas to the specified locations one by one by helicopter. 
Then the height and the direction of the wireless LAN antenna of each SD unit were adjusted to yield the best communication with their corresponding towers.
Data-taking at each communication tower began at the end of April 2019.

\section{Calibration}
\label{Calibration}
\begin{figure}
      \centering
       \includegraphics[width=0.6\textwidth]{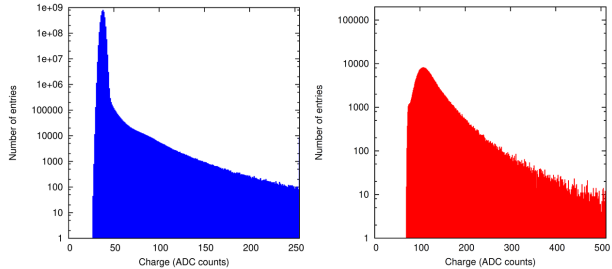}
      \caption{
        Left figure: A typical ADC distribution for the pedestal.
        The distribution was generated by integrating ADC values over 160 ns (20 ns $\times$ 8 bins) that were buffered for ten minutes.
        The total number of events in the histogram is 
        3.75$\times10^9$.
        The histogram contains particle signals, which make up the tail.
        Right figure: A sample of a typical ADC distribution for calibration with single muons.
        The distribution was generated by integrating ADC values over 240 ns (20 ns $\times$ 12 bins) that were buffered for ten minutes.
        For the coincidence signals, the data are taken when both integrated ADC values are greater than the trigger threshold (15 ADC values above the baseline). 
        The total number of events in the histogram is 
        4.45$\times10^5$.
        Histograms of all surface detectors are collected every ten minutes.
      }
      \label{ADC} 
\end{figure}

Atmospheric muons provide a stable source of calibration for the SD units.
We simulated low energy air showers using CORSIKA~\cite{CORSIKA} and their energy deposition in the TAx4 SD scintillators using GEANT4~\cite{GEANT41, GEANT42}. The single-muon peak of the energy deposition in the plastic scintillators of the TAx4 SDs was approximately 2.4 MeV. 
A 12-bit flash analog-to-digital converter (ADC) of the electronics of the TA and TAx4 SDs digitizes waveforms from each PMT.
The flash ADC counts are sampled every 20 ns, and the flash ADC counts in continuous twelve time slices are integrated to obtain single-muon peaks.
A low-pass filter is used for shaping the analog signal, and the -3 dB cutoff frequency of the low-pass filter is 9.7 MHz.
Fig.~\ref{ADC} shows a typical example of ADC distributions of the pedestal and coincidence signals of the two layers of scintillators.
The right figure of Fig.~\ref{ADC} shows a clear single-muon peak. 

\begin{figure}
  \centering
  \includegraphics[width=.5\textwidth]{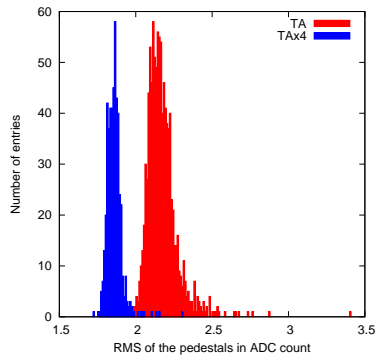}
  \caption{
    The distributions of the RMS width obtained from the Gaussian fit to the pedestal
    distributions (see left figure of Fig.~\ref{ADC}) of the TA and TAx4 surface detectors (SDs).
    The mean RMS width 
    for the TA SDs is $2.16$ ADC counts.
    The mean RMS width 
    for the TAx4 SDs is $1.85$ ADC counts.
    (For comparison, the RMS width of the example pedestal distribution in Fig.~\ref{ADC} is 1.87 ADC counts.) 
    All the pedestal distributions of the TA SDs (2 layers $\times$ 507 SDs $=$ 1014) are included in this figure.
    Most of the TAx4 SDs (2 layers $\times$ 241 SDs $=$ 482) are also included, but some TAx4 SDs now being repaired are not.
    The pedestal values of the TA SDs in this figure were collected in 2009.
  }
  \label{PedestalSigma} 
\end{figure}

We modified the electronics of the TA SDs 
to improve the stability
of the ADC baseline (which otherwise sometimes shifted by a few ADC counts at a specific temperature depending on the circuitry) and 
made the same modification on the TAx4 SDs. 
The pedestal distributions of the TA and TAx4 SDs were fit with Gaussian distributions. 
Fig.~\ref{PedestalSigma} shows the distribution of 
the RMS widths of these Gaussians; for the TAx4 SD, the RMS values are on average 15\% smaller than for the TA SD. 
The difference of the RMS values corresponds to less than 1\% of a single-muon peak, and the influence on the fluctuation of a single-muon peak is small.
The difference mainly affects the fluctuation of the Level-2 trigger rate which is shown in Section~\ref{DAQ}.

The uncertainty in the single-muon peak reflects primarily statistical fluctuations in the number of photoelectrons from the photocathode of the PMT, (i.e., in the number of photons collected by the PMT, multiplied by the quantum efficiency and the collection efficiency~\cite{PMT}). 
We measured the number of photoelectrons corresponding to the single-muon peak for each layer of the SD 
with the high voltage (HV) corresponding to a gain of $5\times10^6$ applied to the PMT\footnote{The relation between HV and a gain of $5\times10^6$ for each PMT was measured by Hamamatsu Photonics.}.
The mean and the RMS width of the number of photoelectrons were 22.6 and 5.4, respectively. The corresponding values for the TA SD were 24.6 and 7.2.
This shows that, 
after the various scintillator-box design-changes described in 
Section~\ref{Design},
the number of photoelectrons corresponding to a single-muon peak is comparable between both SDs.

We also measured the light yield of the plastic scintillators to understand the primary source of variations in the number of photoelectrons. A TAx4 SD is composed of sixteen plastic scintillator sheets. 
First, two plastic scintillator sheets were prepared as a standard reference.  The light from the sheets was read out using WLS fibers, a PMT, and SD electronics. The two sheets were placed on top of each other to emulate the two layers, and the single-muon peaks of the coincidence signals of the two layers were measured.
Choosing one sheet out of every batch of 50 manufactured sheets and replacing the upper reference scintillator with that sheet, we measured the single-muon peaks of the coincidence signals and interpreted the relative peak position differences as the relative differences of the light yields of the sheets.
We repeated the same measurements for the sheets of all TAx4 SD units before assembling them.
We measured the number of photoelectrons corresponding to the single-muon peaks of the TAx4 SDs after we assembled the TAx4 SDs and took the correlation coefficient of the number of photoelectrons with the light yield of the plastic scintillators. The correlation coefficient was 0.92. 

\begin{figure}
  \centering
  \includegraphics[width=.5\textwidth]{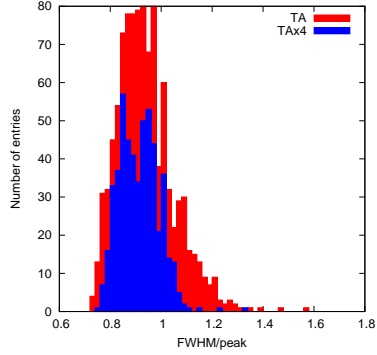}
  \caption{
    The distributions of the FWHM/peak
    of the ADC distributions (see right figure of Fig.~\ref{ADC}) of the TA and TAx4 surface detectors (SDs).
    The mean FWHM/peak of the TA SDs is $0.94$, with the RMS width $0.12$.
    The mean FWHM/peak of the TAx4 SDs is $0.91$, with the RMS width $0.08$.
    All the TA and TAx4 SDs are included in the histograms.
    The total number of entries of the TA SDs is 2 layers $\times$ 507 SDs
    $=$ 1014; that of the TAx4 SDs is 2 layers $\times$ 257 SDs
    $=$ 514.
  }
  \label{FWHMPeak} 
\end{figure}

After the deployment of the SDs, the single-muon peak was adjusted to 
approximately 50 ADC counts to equalize the signal response of the detectors. 
Fig.~\ref{FWHMPeak} shows the distribution of the full width at half maximum (FWHM) of the ADC distribution of coincidence signals, divided by the single-muon peak of the same ADC distribution (FWHM/peak). 
This is a direct measure of the fluctuation of the peak. 
Fig.~\ref{FWHMPeak} uses the single-muon peaks that were obtained from the deployed SDs when the temperature was 
10$^\circ$C.
Because there is no TAx4 SD with a FWHM/peak outside of the distribution of the TA SDs, we can operate the TAx4 SDs in the same way as the TA SDs.
The mean of FWHM/peak for the TAx4 SD is 0.91, and the mean of FWHM/peak for the TA SD is 0.94.

\begin{figure}
      \centering
      \includegraphics[width=0.4\textwidth]{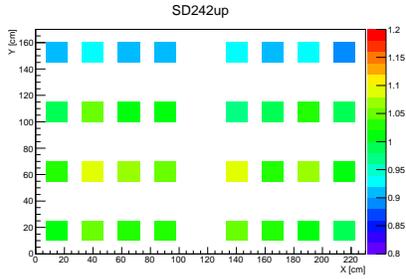}
      \caption{The position dependence of the single-muon peaks 
      for the upper layer of a typical surface detector.
      The x-axis and y-axis denote the dimensions of the 
      scintillator box (see Fig.~\ref{ScintiBox}) parallel and perpendicular to the direction of fibers attached on the scintillators, respectively.
      Colors represent the peak ADC values relative to the mean of the peaks for the upper layer of the detector.
        }  
      \label{SD242up} 
\end{figure}

\begin{figure}
      \centering
      \includegraphics[width=0.4\textwidth]{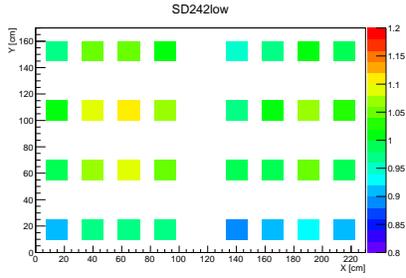}
      \caption{The position dependence of the single-muon peaks of the lower layer of the same surface detector shown in Fig.~\ref{SD242up}.}
      \label{SD242low} 
\end{figure}

We 
measured the position dependence of the single-muon peaks with trigger probes placed on the scintillator box of arbitrarily selected seven TAx4 SDs. 
We placed multiple probes on the box to take data from approximately a quarter of each detector's area at a time. We recorded coincidence signals of the two PMTs of each SD when there is a trigger from one probe PMT.
Fig.~\ref{SD242up} shows the position dependence of 32 single-muon peaks relative to the mean of 32 peaks in the upper layer of a representative TAx4 SD, and Fig.~\ref{SD242low} shows that in the lower layer.
The dependence of the single-muon peaks on the y-axis shows that the minimum values are located at the edge of the y-axis where WLS fibers are curved with the smallest radius of curvature to lay them into the grooves on the scintillator surface, for example as shown at the foreground edge of the upper layer in Fig.~\ref{ScintiBox}.
The single-muon peaks of the upper layer are the smallest at largest y values, and those of the lower layer are the smallest at the smallest y values.
All seven TAx4 SDs
show the same dependence of the single-muon peaks on the y-axis.
This can be understood as a result of the attenuation length of the WLS fibers, which have opposite arrangements in the upper and lower layers. 
The mean of the RMS widths of 
the distributions of the single-muon peaks relative to the mean peaks of the corresponding SDs for all seven TAx4 SDs
is 6\%. 
In all, 94\% of the data points are less than 10\% off from the mean value; the maximum difference from the mean value is less than 20\%.
The corresponding mean of the RMS widths for the TA SDs is 9\%. 
Although the WLS fibers used for the TAx4 SDs are longer than those used for the TA SDs, the dependence of the single-muon peaks on position for the TAx4 SDs is smaller than that for the TA SDs because the position dependence of the photon sensitivity on the PMT surface for the TAx4 SDs is smaller than that for the TA SDs.

\begin{figure}
      \centering
      \includegraphics[width=0.5\textwidth]{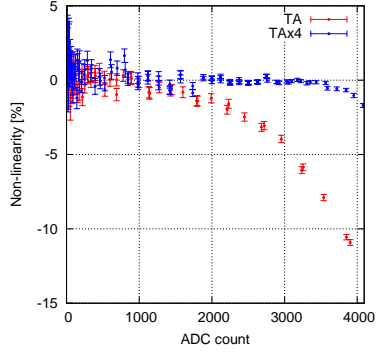}
      \caption{
      Non-linearity of output currents of typical photomultiplier tubes for the TA and TAx4 surface detectors, measured using the LEDs in the detector. Each ADC count corresponds to 0.01 mA. }
      \label{Linearity} 
\end{figure}

\begin{figure}
      \centering
      \includegraphics[width=0.5\textwidth]{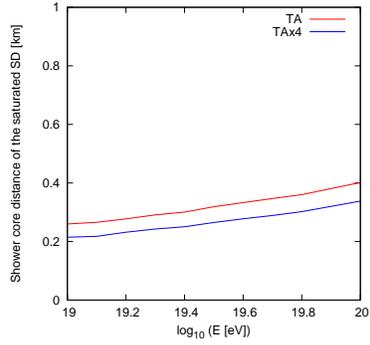}
      \caption{
       	Simulated mean shower-core distances of the saturated TA and TAx4 surface detectors (SDs) as a function of 
       	the energies of the primary cosmic-ray protons.
        The detailed settings of the Monte Carlo simulations are the same as in Fig.~\ref{TrigEff}.
        If the magnitude of the non-linearity defined in Eq.~\ref{non-linearity} is greater than 5\%, the SD is counted as saturated.
      }
      \label{Saturation} 
\end{figure}

We measured the linearity of the magnitude of the signals from the two PMTs after assembling each SD.
Four LEDs are included in each SD: 
two for calibrating PMT linearity for the upper layer, and two for the lower layer.
Pulse linearity of all the PMTs was measured with the LEDs in the same way 
as it had been done for the TA SDs.
We first measured the pulse height (A) of the PMT signal by flashing only one LED of one layer of one SD and measured the pulse height (B) of the PMT signal by flashing only the other LED.
Then we measured the pulse height (C) of the PMT signal by flashing two LEDs simultaneously. In the absence of a non-linearity, one would expect ${\rm C = A + B}$. The non-linearity is defined:
\begin{equation}
\rm \frac{(C)-\{(A)+(B)\}}{(A)+(B)}.
\label{non-linearity}
\end{equation}
The pulse height was measured in ADC counts using the flash ADCs every 20 ns.
Fig.~\ref{Linearity} shows typical measurement results for the TA and TAx4 SDs, obtained using the electronics for the SDs. For the TA SD, the PMT non-linearity is within 5\% in the range from 0 to 2930 ADC counts.  For the TAx4 SD, it is within 5\% for the entire range from 0 to 4095 ADC counts. 
When the non-linearity is 5\%, the mean pulse height of the PMTs for the TA SDs is 2900 ADC counts, corresponding to 29 mA.
When the signals of the PMTs are not linear, the signals are excluded from the fitting of the lateral distribution in the event reconstruction. We used air-shower simulations~\cite{TASDMC} to estimate the core distances of the non-linear PMT signals. 
Fig.~\ref{Saturation} depicts simulated core distances when the non-linearity of the signals is greater than 5\% as a function of the primary energies of the cosmic rays. The mean of the core distances of the saturated TAx4 SDs is between 210 m and 340 m when the primary energies are greater than 10 EeV, whereas the mean of the core distances of the saturated TA SDs is expected to be between 260 m and 400 m under the same conditions.

The SD electronics is running with a 50 MHz internal clock. The clock rate is not exactly 50 MHz and gradually changes depending on the temperature.
Therefore, the 50 MHz internal clock of each SD is calibrated using a 1 pulse-per-second (PPS) signal from the GPS receiver (M12M i-Lotus) included in the SD electronics. 
We checked the accuracy of the calibration at the test bench in the following manner:
One test pulse from a function generator was divided equally, and 
the divided pulses were input to several SDs.
The signal of the GPS antenna was also divided, and identical signals were input to the GPS receivers in the SD electronics. 
A specific board of the SD electronics and a GPS receiver with the serial number RD4569 were used as a standard reference for this test.
The difference of the trigger 
time
between the SD electronics and the standard reference was measured.  
Since this offset in the trigger time is caused only by the GPS receivers, we repeated the measurement for all of the GPS receivers to estimate their time resolution.

\begin{figure}
      \centering
      \includegraphics[width=0.5\textwidth]{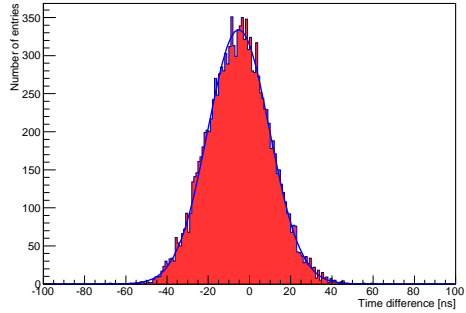}
      \caption{ 
        A typical sample of the distribution of the trigger timings of one circuit board of an SD electronics with respect to the standard reference electronics. 
        The mean of this distribution is -5.3 ns, and the RMS width is 15.1 ns; 
        the blue curve is a Gaussian fit to the distribution. 
        There are 102 degrees of freedom; $\chi^2 = $123. The total number of entries in this histogram is 12631.
        }
      \label{GPS} 
\end{figure}

\begin{figure}
      \centering
      \includegraphics[width=0.5\textwidth]{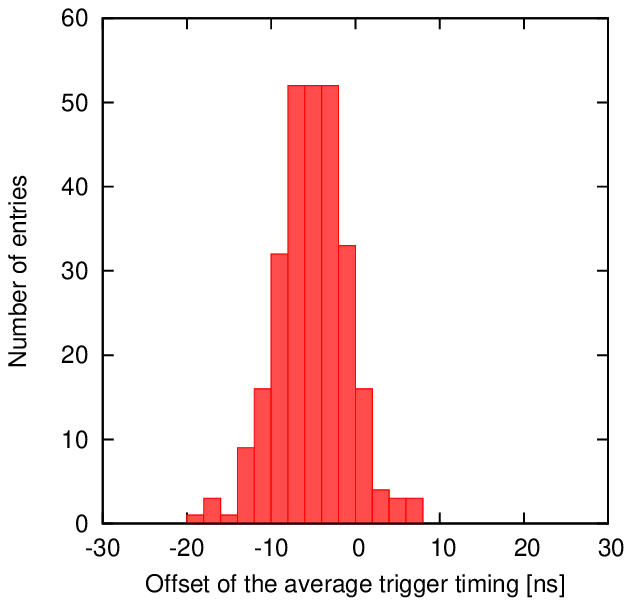}
      \caption{ 
        The distribution of the mean of the relative trigger timings (see Fig.\ref{GPS}) of 374 different GPS receivers. 
        The mean of this distribution is -3.6 ns, and the RMS width is 5.1 ns. 
        The largest difference from the standard GPS-receiver value is -19.3 ns.
        }
      \label{GPS_offset} 
\end{figure}

A typical distribution of the difference of the trigger timings is shown in Fig.~\ref{GPS}.
We fitted each distribution to a Gaussian.
The mean of the RMS widths of the Gaussian distributions of all the measurements was 14.9 ns, so the resolution of the trigger timing of one GPS receiver can be estimated
as $14.9/\sqrt{2} = 10.5$ ns on average. 
The offsets of the trigger timings are shown in Fig.~\ref{GPS_offset}. 
The mean offset from the standard reference was -3.6 ns and the RMS width of the offset distribution was 5.1 ns. The largest offset was -19.3 ns.
The systematic uncertainty of the determination of the arrival directions caused by this offset is expected to be well within 0.2 degrees. 
This is much smaller than the angular resolution of the TAx4 SDs of 2.2 degrees 
and therefore, it is adequate for the TAx4 SD event reconstruction.
The 
time
offset of the GPS receiver does not need to be adjusted by the firmware of the electronics.

\begin{figure}
      \centering
      \includegraphics[width=0.5\textwidth]{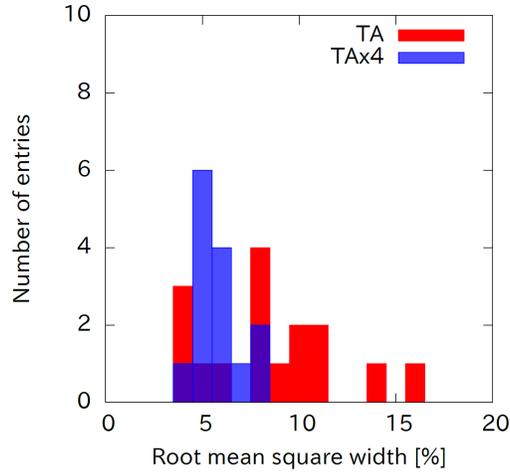}
      \caption{The distribution of the RMS width of the position dependence of the single-muon peaks relative to the mean peaks. This figure includes the position dependence of eight TA surface detectors (SDs) and seven TAx4 SDs. The position dependence of a TAx4 SD is shown in Fig.~\ref{SD242up} and Fig.~\ref{SD242low}.}   
      \label{RMSwidth} 
\end{figure}

\section{Data acquisition}
\label{DAQ}

There are three types of triggers (Level-0, Level-1, and Level-2) in both the TA and TAx4 data-acquisition systems. 
The SD electronics collects signals and converts them to ADC values every 20 ns, using a 12-bit flash ADC, to obtain their waveforms. 
The Level-0 trigger is used to record the waveforms into the buffer of the SD electronics.
Its threshold is 15 ADC counts, which corresponds to approximately 1/3 of the single-muon peak, above the baseline.
The data size of the Level-0 trigger 
times
is too large to send, so the Level-1 trigger, with a threshold of 150 ADC counts above the baseline, is used instead.
The Level-1 trigger 
time stamps
are sent to the communication tower for the evaluation by the Level-2 trigger, which corresponds to the trigger to collect an air-shower event as described below, which in turn is judged by the electronics at the communication tower. 
The tower electronics searches for three adjacent Level-1 (over 150 ADC counts) SDs triggered within 14 $\mu$s of one another.
The only difference between the triggers in the TA SDs and TAx4 SDs is this gate width, which is 8 $\mu$s in the case of the TA SDs.
When the collected Level-1 triggers match the above condition, the Level-2 trigger is issued from the communication tower to the SDs in the sub-array of the corresponding tower.
(The trigger condition was reflected in the simulation in Fig.~\ref{TrigEff} in Section \ref{Design}.)
All the recorded waveforms using the Level-0 trigger are collected from every SD in the sub-array within 
$\pm$32 $\mu$s of the trigger 
time
once a Level-2 trigger is generated.

The electronics of each TAx4 SD evaluates the Level-0 and Level-1 triggers and sends the trigger 
times
and waveforms to the communication tower.
The design of the electronics is the same as in the TA SD except for the wireless LAN module and the following two modifications. 
We modified the circuit 
to produce a more stable ADC baseline.
We also added diodes, thereby avoiding 
excessively large signals;
this was necessary to use for the TAx4 SD PMTs. 

\begin{figure}
      \centering
      \includegraphics[width=0.5\textwidth]{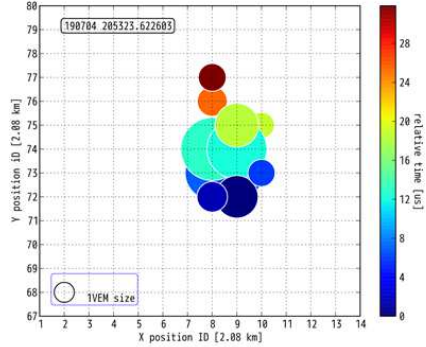}
      \caption{ 
        The footprint of a detected air-shower event. 
        The labels for the x-axis and y-axis represent the position IDs of the surface detectors in the east-west and the north-south directions, respectively.
        Circle area corresponds to the logarithmic size of the signal. Color represents the arrival time at the detector.
      }
      \label{EventDisplay_foot} 
\end{figure}

\begin{figure}
      \centering
      \includegraphics[width=0.5\textwidth]{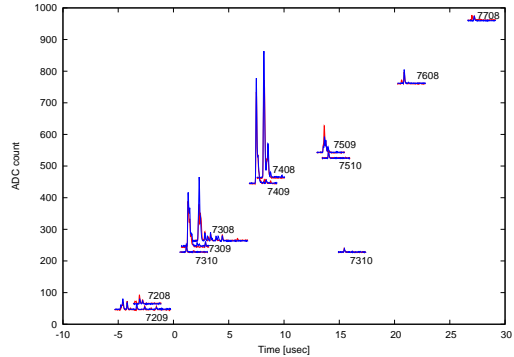}
      \caption{ 
        Flash ADC waveforms of 
        the detectors of the event shown
        in Fig.~\ref{EventDisplay_foot}.
        The red lines show waveforms of the signals of the upper scintillators, and the blue lines show waveforms of the signals of the lower scintillators.
ADC values of each surface detector counter are shifted by adding offsets that are proportional to the counter distance from the shower axis.
        Four-digit numbers are the position IDs of surface detectors: 
the first two digits are the position ID along the y-axis, and the last two digits are the position ID along the x-axis (see Fig.~\ref{EventDisplay_foot}).
      }
      \label{EventDisplay_wf} 
\end{figure}

The wireless communication protocol was also changed from the customized protocol of the TA SDs to the user datagram protocol (UDP), permitting us to use the new wireless LAN module (WVCWB-R-022(05) WiViCom) because the module used for the TA SD was out of production. 
The transmission control protocol (TCP) was considered also. However, UDP, unlike TCP, does not require established socket connections prior to transmitting network packets, thus allowing a higher data collection rate.
On the other hand, UDP does not check whether the data packets were received or not. Therefore, a retry process was manually introduced in the data acquisition program. 
The retry process reduced the packet loss rate 
by a factor of $\sim$10
in the data taking using a test SD array, and the packet loss rate was approximately 0.01\%.
A single-board computer (Raspberry Pi 2 model B) is used for generating the Level-2 triggers and for taking data from the SDs at each communication tower.
The single-board computer communicates with the SDs via an access point 
(Aironet AP1572E-Q-K9 Cisco) established at the
communication tower~\cite{comtower}.
Fig.~\ref{EventDisplay_foot} and Fig.~\ref{EventDisplay_wf} show, respectively, the footprint and the SD waveforms of an air shower event that was recorded using this TAx4 SD data acquisition system.
This new wireless communication system was first introduced for the TA low energy extension (TALE) experiment~\cite{TALE}, and it has been successfully used by the TALE surface detector.

\section{Summary}
\label{FuturePros}
TA saw evidence of anistropy in the distribution of cosmic rays with energies greater than 57 EeV. 
To collect data connected to this possible discovery more rapidly, we have begun construction of a new larger detector array, TAx4, northeast and southeast of the TA SD array. In TAx4, the SDs are arranged on a square grid with a 2.08 km spacing.
On the basis of simulations, we expect 25\% energy resolution, 2.2$^\circ$ angular resolution, and 95\% reconstruction efficiency of cosmic rays with energies greater than 57 EeV.

We made several important design changes to the interior of the scintillator boxes, such as using 
PMTs with higher quantum efficiency and more uniform photon sensitivity on the surface and reducing the total length of the WLS fibers by 67\%.
Since there were other changes in addition to the changes described above for PMTs and fibers, as a result,
the effect was not so large
on the distribution of the FWHM/peak or position dependence of the single-muon peaks, and
the PMTs of the TAx4 SDs have a wider range linearity than those of the TA SDs.
The mean of the core distances of the saturated TAx4 SDs is expected to be between 210 m and 340 m when the primary energies are greater than 10 EeV, whereas the mean of the core distances of the saturated TA SDs is expected to be between 260 m and 400 m under the same conditions.

We modified the electronics for the TAx4 SDs.
The mean RMS width of their pedestals is 15\% smaller than that 
of the pedestals of the TA SDs. The baseline of some of the TA SD electronics was observed to shift by a few ADC counts at certain temperatures; this problem was solved by the modification. 

We measured the time offsets of all GPS receivers. 
The mean time offset with respect to the standard reference was -3.6 ns, and the RMS width was 5.1 ns. The largest time offset was -19.3 ns.
The systematic uncertainty of the determination of the arrival directions caused by this offset is expected to be less than 
0.2$^\circ$, which is  small enough for the requirement of the TAx4 SD shower reconstruction.

The UDP communication protocol was introduced for the TAx4 SDs, and used to realize a data-acquisition system similar to that of the TALE SDs.
The only difference between the trigger conditions of the TA and TAx4 SDs is the gate width of the Level-2 trigger: 14 $\mu$s for the TAx4 SDs and 8 $\mu$s for the TA SDs.

We deployed 257 TAx4 SDs in February and March 2019 and started 
collecting data using the entire TAx4 SD array at the end of April 2019.

\section*{Acknowledgements}

The Telescope Array experiment is supported by the Japan Society for
the Promotion of Science(JSPS) through 
Grants-in-Aid
for Priority Area
431,
for Specially Promoted Research 
JP21000002, 
for Scientific  Research (S) 
JP19104006, 
for Specially Promoted Research 
JP15H05693, 
for Scientific  Research (S)
JP15H05741, for Science Research (A) JP18H03705, 
for Young Scientists (A)
JPH26707011, 
and for Fostering Joint International Research (B)
JP19KK0074,
by the joint research program of the Institute for Cosmic Ray Research (ICRR), The University of Tokyo; 
by the Pioneering Program of RIKEN for Evolution of Matter in the Universe (r-EMU);
by the U.S. National Science
Foundation awards PHY-1404495, PHY-1404502, PHY-1607727, PHY-1712517, and PHY-1806797; 
by the National Research Foundation of Korea
(2017K1A4A3015188, 2020R1A2C1008230, \& 2020R1A2C2102800) ;
by the Ministry of Science and Higher Education of the Russian Federation under the contract 075-15-2020-778, RFBR grant 20-02-00625a (INR), IISN project No. 4.4501.18, and Belgian Science Policy under IUAP VII/37 (ULB). The foundations of Dr. Ezekiel R. and Edna Wattis Dumke, Willard L. Eccles, and George S. and Dolores Dor\'e Eccles all helped with generous donations. The State of Utah supported the project through its Economic Development Board, and the University of Utah through the Office of the Vice President for Research. The experimental site became available through the cooperation of the Utah School and Institutional Trust Lands Administration (SITLA), U.S. Bureau of Land Management (BLM), and the U.S. Air Force. We appreciate the assistance of the State of Utah and Fillmore offices of the BLM in crafting the Plan of Development for the site.  Patrick A. Shea assisted the collaboration with valuable advice and supported the collaboration’s efforts. The people and the officials of Millard County, Utah have been a source of steadfast and warm support for our work which we greatly appreciate. We are indebted to the Millard County Road Department for their efforts to maintain and clear the roads which get us to our sites. We gratefully acknowledge the contribution from the technical staffs of our home institutions. An allocation of computer time from the Center for High Performance Computing at the University of Utah is gratefully acknowledged.






\end{document}